\documentclass[preprint,prb,twocolumn,10pt,tightenlines,oneside]{revtex4}%
\usepackage{amsfonts}
\usepackage{amsmath}
\usepackage{amssymb}
\usepackage{graphicx}%
\providecommand{\U}[1]{\protect\rule{.1in}{.1in}}

\begin{document}
\preprint{ }
\title[Electrostatic and Parallel Magnetic Field Tuning]{Electrostatic- and Parallel Magnetic Field- Tuned Two Dimensional
Superconductor-Insulator Transitions}
\author{Kevin A. Parendo, K. H. Sarwa B. Tan, and A. M. Goldman}
\affiliation{School of Physics and Astronomy, University of Minnesota, Minneapolis, MN
55455, USA}
\pacs{PACS number}

\begin{abstract}
The two-dimensional superconductor-insulator transition in disordered
ultrathin amorphous bismuth films has been tuned both by electrostatic
electron doping using the electric field effect and by the application of
parallel magnetic fields.\ Electrostatic doping was carried out in both zero
and nonzero magnetic fields, and magnetic tuning was conducted at multiple
strengths of electrostatically induced superconductivity. The various
transitions were analyzed using finite size scaling to determine their
critical exponent products. For the electrostatically tuned transition the
exponent product $\nu z=0.7\pm0.1$, using data from intermediate temperatures
down to 60 mK. Here $\nu$\ is the correlation length exponent and $z$ is the
dynamical critical exponent.\ In the case of electrostatically tuned
transitions in field, and the field-tuned transtions at various values of
electrostatically induced superconductivity, from intermediate temperatures
down to about 100 to 150 mK scaling was successful with $\nu z=0.65\pm0.1$.
The parallel critical magnetic field, B$_{c}$, increased with electron
transfer as $(\Delta n_{c}-\Delta n)^{0.33}$, and the critical resistance
decreased linearly with $\Delta n$. However at lower temperatures, in the
insulating regime, the resistance became larger than expected from
extrapolation of its temperature dependence at higher temperatures, and
scaling failed. These observations imply that although the electrostatic- and
parallel magnetic field- tuned superconductor-insulator transitions would
appear to belong to the same universality class and to be delineated by a
robust phase boundary that can be crossed either by tuning $\Delta n$ or $B$,
in the case of the field-tuned transition at the lowest temperatures, some
different type of physical behavior turns on in the insulating regime.

\end{abstract}
\volumeyear{year}
\volumenumber{number}
\issuenumber{number}
\eid{identifier}
\startpage{1}
\endpage{ }
\maketitle

\section{Introduction}

Continuous quantum phase transitions are transitions at absolute zero in which
the ground state of a system is changed by varying a parameter of the
Hamiltonian. \cite{Sondhi, Sachdev}\ The transitions between superconducting
and insulating behavior in two-dimensional superconductors tuned by magnetic
field or disorder (thickness) are believed to be such transitions. Early
experiments and theories seemed to support a picture of only two ground
states. \ Historically the first theoretical approach to the
superconductor-insulator (SI) transition was based on Cooper pairing being
suppressed by the enhancement of the Coulomb repulsion between electrons with
increasing disorder. \cite{Fukuyama, Belitz, Finkelshtein} In effect the order
parameter would be suppressed to zero in the insulating regime. \ The absence
of a gap in the density of states in tunneling studies of the insulating
regime has been interpreted as evidence for a zero superconducting order
parameter amplitude in the insulating regime. \cite{Dynes and Valles} Another
approach to the SI transition was based on the transition being governed by
phase fluctuations. \cite{FWGF, Fisher, FGG} \ In this \textquotedblleft dirty
Boson model,\textquotedblright\ the insulator is a vortex condensate with
localized Cooper pairs, in contrast with the superconductor, which is a Cooper
pair condensate with localized (pinned) vortices. Elaborations on this model
have included effects of fermionic degrees of freedom. \cite{Trivedi}\ If
phase fluctuations are pair breaking and destroy the superconducting energy
gap, then this picture would also be consistent with the tunneling studies.

The three conventional approaches to the tuning of SI transitions all involve
uncontrolled aspects of morphology. In the first method, a relatively thick
($\backsim100$\AA \ ) superconducting film is produced and magnetic fields are
applied to quench the superconductivity.\ These films are two dimensional in
the sense that the coherence length and penetration depth are much greater
than the films' thicknesses.\ In this approach, the strength of the vortex
pinning, which depends on the nature of the disorder, may determine the
outcome of the measurements.\ In some measurements on films with relatively
low sheet resistances \cite{Yazdani and Kapitulnik}, the transition can also
be described using quantum corrections to conductivity \cite{Larkin and
Galitski} and may not be quantum critical at all. \cite{Baturina} In a second
approach, the thickness of a film is increased in small increments, tuning
from insulator to superconductor. \cite{Jaeger, Haviland} However, films of
different thicknesses may have different morphologies. \ In a third method,
which has been used in some studies of In$_{2}$O$_{3}$ films, an insulating
film is thermally annealed to produce superconductivity. \cite{Ovadyahu,
Gantmakher anneal}\ However, thermal annealing may involve the alteration of
morphology and chemical composition.

Successful finite-size scaling analyses with film thickness or magnetic field
as tuning parameters have resulted in critical exponent products, $\nu z$, in
the range of 1.2 to 1.4, \cite{Hebard and Paalanen, Nina 2D SIT, Markovic
Phase, Yazdani and Kapitulnik, Gantmakher scaling} which have been suggested
to result from the transition being dominated by percolative effects,
\cite{Shimshoni and Kapitulnik, Meir} as this number is close to the exponent
in 2D percolation.\ One investigation of a perpendicular field tuned
transition has yielded 0.7 as the product.\cite{Nina 2D SIT}

The simple two-ground state picture has been challenged in recent work, which
appears to indicate that there is an extended intermediate metallic regime
over an extended range of the tuning parameter. The physical evidence for such
a regime is that resistances become independent of temperature at the lowest
temperatures both in perpendicular field tuned transitions\ \cite{Mason and
Kapitulnik} and in thickness-tuned transitions. \cite{Parendo PRB} In the case
of the magnetic field experiments, the metallic regime abruptly disappears as
the magnetic field is reduced, leading to true superconductivity.\ This is
further evidence of possible sample inhomogeneity.\ There are several theories
that describe an intermediate metallic regime \cite{Fisher 2005, Lopatin,
Phillips} in homogeneous samples.\ However it is not established that this
regime is indeed intrinsic, and is not a consequence of either sample
inhomogeneity, or the failure to cool a film.\ In overtly granular systems
there is a clear intermediate metallic regime that is found at temperatures in
excess of 1 K that is probably intrinsic.\cite{Jaeger, Christiansen}

Another phenomenon reported by a number of groups is the appearance of a large
peak in the resistance at fields in excess of the critical field for the SI
transition for superconducting amorphous In$_{2}$O$_{3}$ and TiN films,
\cite{Paal Heb Ruel, Gantmakher metal, Gantmakher quantum corrections,
Villegier, Shahar, Steiner and Kapitulnik, Wang, Baturina} or intrinsically
insulating amorphous Be films.\cite{Butko and Adams}\ With the exception of
the work of Gantmakher \textit{et al}., \cite{Gantmakher parallel} most of
these studies have been conducted in perpendicular magnetic fields No
quantitative theoretical explanation has yet been put forward to explain these
very large values of resistance found on the insulating side of the SI
transition. It is not clear that this phenomenon is an intrinsic property of a
homogeneous material or results from some mesoscale inhomogeneity. \ 

Since there are important concerns about morphology and disorder in the
above-mentioned studies, the aim of the present work was to attempt to clarify
these issues by studying SI transitions in which the level of physical
disorder was fixed, and in which the outcome was not dependent upon the degree
of vortex pinning. This is possible by inducing superconductivity in an
insulator by electrostatic doping using the electric field effect
\cite{Parendo PRL} and then applying a parallel magnetic field to drive the
film back into the insulating state.\ The same level of chemical and physical
disorder may be shared by both the intrinsic insulating state, and the
insulating state in which the electrostatically induced superconductivity is
quenched by magnetic field.\ We will present arguments to the effect that
electrostatic doping does not alter physical or chemical disorder, but changes
the coupling constant that determines superconductivity.\ Parallel magnetic
fields destroy superconductivity by polarizing spins, but not by inducing
vortices.\ These studies were carried out in a very carefully shielded
dilution refrigerator to enhance the chances that very low heat capacity films
would cool.

The experimental approach used in this work will be described in Section II.
\ In Section III the various results will be presented.\ The final Section
contains a discussion of these results along with conclusions that can be
drawn. \ 

\section{Experimental Approach}

These investigations were carried out in a geometry in which a SrTiO$_{3}$
(STO) crystal served as both a substrate and a gate insulator in a field
effect transistor configuration.\ To prepare this device, first a small
section of the unpolished back surface of a 500$\mu$m thick single-crystal of
(100) STO substrate was mechanically thinned\cite{Bhattacharya}
\textit{ex-situ}, resulting in this surface and the epi-polished front surface
being parallel and separated by $45\pm5$ $\mu$m. \ A 0.5 mm by 0.5 mm, 1000
\AA \ \ thick, Pt \textquotedblleft gate\textquotedblright\ electrode was
deposited \textit{ex situ} onto the thinned section of the back surface
directly opposite the eventual location of the measured square of film
Platinum electrodes, 100 \AA \ \ thick, were also deposited \textit{ex situ}
onto the substrate's epi-polished front surface to form a four probe
measurement geometry.\ The substrate was then placed in a Kelvinox-400
dilution refrigerator/UHV deposition apparatus. \cite{Hernandez RSI} \ A 10
\AA \ thick under-layer of amorphous Sb and successive layers of amorphous Bi
(\textit{a-}Bi) were thermally deposited \textit{in situ} under ultra-high
vacuum conditions ($\backsim10^{-9}$ Torr) through shadow masks onto the
substrate's front surface.\ The substrate was held at about 7 K during the
deposition process.\ Films grown in this manner are believed to be
homogeneously disordered on a microscopic, rather than on a mesoscopic
scale.\cite{Strongin}

A Keithley 487 voltage source was used to apply voltages between the film and
the gate electrode.\ In essence, the film and the gate electrode formed a
parallel plate capacitor with the thinned layer of STO serving as the
dielectric spacer. Applying a positive voltage, $V_{G}$, to the gate electrode
caused electrons to be transferred into the film. \ Since STO crystals have
very large dielectric constants at temperatures below 10 K ( $\backsim
20,000$), and since the substrate was greatly reduced in thickness, the gate
voltage produced large electric fields that facilitated large transfers of
electrons.\ Since the dielectric constant of STO is known to vary strongly
with electric field, an analysis was performed that yielded the relationship
between $V_{G}$ and the areal density of added electrons, $\Delta n$, for this
substrate and film.\ At positive gate voltages, transferred electron densities
were found to be between 0 (at $V_{G}=0$) and $3.35$ x $10^{13}$ cm$^{-2}$ (at
$V_{G}=42.5$ V).\ Application of $V_{G}$ above 42.5 V did not change any
measurement of film properties relative to those at 42.5 V.

The sample measurement lines were heavily filtered so as to minimize the
electromagnetic noise environment of the film. The approach was to use RC
filters at 300 K to attenuate 60 Hz noise, Spectrum Control \#1216-001 $\pi
-$section filters at 300 K (in series with 10 k$\Omega$ resistors) to
attenuate radio frequency noise, and 2 m long Thermocoax cables \cite{Zorin}
at the mixing chamber stage of the refrigerator to attenuate GHz Johnson noise
from warmer parts of the refrigerator.\ To avoid complications arising from
this filtering, measurements were made using DC, rather than AC, methods. \ A
1 nA DC current, I, was applied to the films using a Keithley 220 current
source. \ Voltage, V, was measured across the centermost 0.5 mm by 0.5 mm
square of the film using a Keithley 182 voltmeter.\ The sheet resistance of
the film, R, was taken to be V/I.

The films clearly fail to cool much below 60 mK, even though the dilution
refrigerator cools to 7 mK. This is almost certainly due to the thermal load
on the films that is caused by the residual noise environment, together with
limitations on the thermal grounding of the electrical leads. It is very
difficult to cool the electrons of a film at such low temperatures. Ultrathin
films have a negligible heat capacity and since in this instance, they are not
immersed in a cooling fluid, the mechanism for the electrons to cool is
through their thermally (but not electrically) grounded leads and through
contact of the film with the \textquotedblleft thermal bath\textquotedblright%
\ of phonons in the film. At mK temperatures, however, electrons and phonons
are known to decouple.\cite{Clarke} The mixing chamber cools to 7 mK, as
verified by a $^{60}$Co nuclear orientation thermometer. To determine the
actual temperature of a film, we used its electrical resistance in the
insulating regime as a thermometer. As will be discussed in Section III, a
good fit to the temperature dependent resistance from 12 K down to 60 mK was
an exponential activation form characteristic of 2D Mott variable range
hopping. However, at 60 mK, R(T) began to deviate from this form and
eventually became independent of temperature as the mixing chamber approached
7 mK. This is consistent with the electrons of the film not cooling even while
the mixing chamber continues to cool.\ Removal of the 60 Hz filtering raised
the temperature at which the 9.60 \AA \ \ thick film began to deviate from the
Mott form at 140 mK; above this temperature, R(T) was relatively unchanged.
This suggests that the noise environment prevented the film from cooling after
the removal of the filter. By an extension of this logic, we believe that,
with the full filtering present, the residual noise environment prevented the
film from cooling below about 60 mK.

Since\textit{ in situ} rotation of the sample into alignment perpendicular to
the axis of the magnetic was not possible during this series of experiments,
Hall effect measurements that could have yielded the intrinsic charge density
were not performed. Electrons were stated by Buckel to be the charge carriers
in \textit{a-}Bi, from a determination of the Hall coefficient in relatively
thick ($\backsim$500 \AA \ ) films.\cite{Buckel} However, the sign of the Hall
coefficient is known to lead to an erroneous determination of the sign of the
charge carrier in amorphous semiconductors.\cite{Kakalios} \ If \textit{a-}Bi
acted like a semiconductor in this regard, then holes would be the charge
carriers, which would make it similar to most other superconductors.
Thermopower measurements could accurately ascertain the sign of the of charge
carriers in \textit{a}-Bi.\cite{Kakalios} Buckel's Hall effect measurements
yield estimates of areal charge densities of between 10$^{14}$ and 10$^{15}$
cm$^{-2}$ in metallic \textit{a-}Bi films. \ 

Careful attention was given to ensure that voltages were measured only while
the film was in thermal equilibrium. Ramping $V_{G}$ even at slow rates caused
heating of the film because real currents flowed in the film in response to
the displacement currents that were induced by changing $V_{G}$. Ramping the
magnetic field caused Eddy current heating.\ Lastly, it was found that the
rate at which the film cooled and warmed in high magnetic fields decreased
dramatically compared to the rate at low fields. Thus, the measurement
procedure that was used involved ramping $V_{G}$ and/or magnetic field to
constant values and waiting long times (between three and six hours) to fully
cool the film to 60 mK. After this, voltages were measured as a function of
temperature from 60 mK to 1 K in controlled steps using temperature feedback
and long equilibration times (between 5 and 15 minutes per temperature).
Appropriate equilibration times after ramping $V_{G}$, $B$, and $T$ were
determined by checking reproducibility, i.e., by taking measurements after
longer wait times, with varying rates, and by recording voltage for many hours
at stable temperatures while watching for drift that would signal behavior
that was out of equilibrium. \ 

In a quantum phase transition, the resistances of 2D films are expected to
obey the finite size scaling functional form\cite{Fisher}
\begin{equation}
R/R_{c}=F\left(  |K-K_{c}|/T^{1/\nu z}\right)  \label{fsc}%
\end{equation}
where $F$ is an unknown function, $K$ is the tuning parameter, $K_{c}$ is the
critical value of the tuning parameter, $R_{c}$ is the critical resistance,
\ is the correlation length critical exponent, and $z$ is the dynamical
critical exponent. To analyze data using this finite size scaling form, we
first plotted isotherms of $R$ $\left(  K\right)  $. \ A crossing point
separates the insulating and superconducting phases, yielding the critical
values $R_{c}$ and $K_{c}$. As temperature becomes too high, subsequent
isotherms will not cross at a single point ($R_{c},K_{c}$), and these
temperatures are then excluded from the analysis. After determining $R_{c}$
and $K_{c}$, all of the parameters characterizing the scaling analysis are
known except for the product $\nu z$. This product is then taken as the
unknown, and $R/R_{c}$ is plotted against $|K-K_{c}|T^{-1/\nu z}$ for various
values of $\nu z$. \ The value of $\nu z$ that produces the best collapse of
the data is then the exponent product.

\section{Results}

\subsection{The Intrinsic Insulating Regime}

A sequence of \textit{a-}Bi films was studied. As a function of increasing
thickness, there were four thicknesses that were insulating and nine that were
superconducting. \ Here we focus on the four insulating thicknesses that we
refer to as \textquotedblleft intrinsic\textquotedblright\ insulators.\ For
the data shown in Fig. 1, there was no gate bias and no applied magnetic
field.%
\begin{figure}
[ptb]
\begin{center}
\includegraphics[
height=2.0695in,
width=3.0199in
]%
{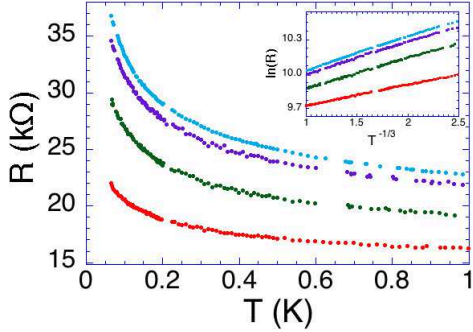}%
\caption{$R(T)$ for thicknesses of 9.60, 9.69, 9.91, and 10.22 \AA \ (top to
bottom in the curves in main figure and inset). Data for $\Delta n=$ 0, $B=$
0, and $T$ from 60 mK to 1 K, shown as $R(T)$ (main) and $lnR$ vs$.T^{-1/3}$
(inset). \ $T_{0}$ and $R_{0}$ decrease with thickness, being 39, 32, 25, and
8 mK and 16406, 16001, 14317, and 13310 $\Omega$ , respectively.}%
\end{center}
\end{figure}
The temperature dependence of the resistance, $R(T)$, at each thickness was
consistent with Mott variable range hopping (VRH) conduction in 2D, as the
best fits to the data were of the form
\begin{equation}
R\left(  T\right)  =R_{0}\exp([T_{0}/T]^{1/3}) \label{mvrh}%
\end{equation}
where the constants $R_{0}$ and $T_{0}$ are the resistance pre-factor and the
activation energy (in K), respectively. \ This is shown in the inset to Fig.
1. \ For the 9.69 \AA \ \ and 9.91 \AA \ thick films, data was available to
verify this form up to 12 K. Unsuccessful attempts were made to fit the data
with hopping powers of 1/3, 1/2, 0.7 and 1 as well as the relationship
expected for weak localization and electron-electron interactions,
\begin{equation}
G\equiv1/R=G_{0}+k\ln(T/T_{0}) \label{wl}%
\end{equation}
where $G$ is the conductivity and $k$ and $T_{0}$ are constants Fits were not
attempted in which the prefactor $R_{0}$ was taken to be temperature
dependent. \ 

Mott VRH conduction is found in strongly localized systems in the absence of
Coulomb interactions between electrons.\ For this work, the absence of Coulomb
interactions may be caused by screening of the electric fields in the
film\cite{Efros and Shklovskii} because of proximity to the STO substrate,
which has a very high dielectric constant.

\subsection{Electrostatic Tuning of the SI Transition in Zero Magnetic Field}

In the case of the 10.22 \AA \ \ thick film, which exhibited the highest
conductivity for the four intrinsically insulating thicknesses of the
sequence, the addition of electrons to the film induced superconductivity. The
evolution of this electrostatically tuned SI transition with charge transfer
is shown in Fig. 2.%
\begin{figure}
[ptb]
\begin{center}
\includegraphics[
height=2.0695in,
width=3.0199in
]%
{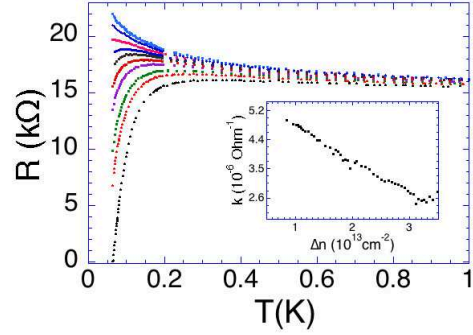}%
\caption{$R(T)$ as a function of $\Delta n$ for the 10.22 \AA \ thick film
with $B=0$. \ Data is shown from 60 mK to 1 K. \ The values of $\Delta n$ that
are shown are 0, 0.62, 1.13, 1.43, 1.61, 1.83, 2.04, 2.37, 2.63, and 3.35 x
10$^{13}$ cm$^{-2}$. \ Forty four curves of $R(T)$ for other values of $\Delta
n$ are omitted from the plot for clarity. Inset: slope of $lnT$ from Eq. 3,
$k$ vs. $\Delta n.$}%
\end{center}
\end{figure}
\ With the charge transfer, $\Delta n=0$, $R(T)$ was well described by Mott
hopping. With $\Delta n=$ 0.85 x 10$^{13}$ cm$^{-2}$, the best fit from 60 mK
to 1 K was that of a $ln(T)$ dependence of the conductance on temperature as
given in Eq. \ref{wl}. This was the first density at which $G(T)$ was
definitely fit better by $ln(T)$ than by the Mott hopping form given in Eq.
\ref{mvrh}. At larger values of $\Delta n$, as superconducting fluctuations
became strong at lower temperatures, this $ln(T)$ dependence remained the best
fit at higher temperatures. With $\Delta n=$3.35 x 10$^{13}$ cm$^{-2}$, the
film was fully superconducting (within the scatter of our data) with the
superconducting transition temperature $T_{c}=$ 60 mK. Superconducting
fluctuations were strong between 60 mK and about 250 mK, and the conductance
was best described by the $ln(T)$ temperature dependence above 250 mK. The
value of the slope of $lnT$ is plotted as a function of $\Delta n$ for
superconducting curves in the inset to Fig. 2. \ The fact that the slope
varies in a linear fashion with $\Delta n$ is striking. The\ actual mechanism
for this $lnT$ dependence is unknown. We have previously suggested that it may
be due to a combination of weak localization and electron-electron
interactions despite the high value of film resistance.\cite{Parendo PRL} \ If
one assumed that only the electron-electron interaction contribution change,
then this would imply that the Hartree screening parameter changed linearly
with $\Delta n$, perhaps contributing to the inducing of superconductivity. \ 

This transition was successfully analyzed using finite size scaling employing
$\Delta n$ as the tuning parameter. This suggests that the electrostatically
tuned SI transition is a quantum phase transition. In the inset to Fig. 3 we
show $R(\Delta n)$ for multiple isotherms between 60 mK and 140 mK. There is a
distinct crossing point at the critical electron density, $\Delta n_{c}=$ 1.28
x 10$^{13}$ cm$^{-2}$ and the critical resistance, $R_{c}=$ $19,109$ $\Omega$
. In the main part of Fig. 3, we show the scaling plot.%
\begin{figure}
[ptb]
\begin{center}
\includegraphics[
height=2.0695in,
width=3.0199in
]%
{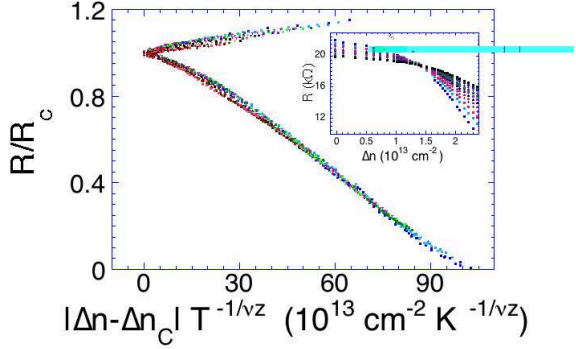}%
\caption{Finite size scaling plot for the 10.22 \AA \ thick film with $B=0$,
including data from 60 mK to 140 mK with $\Delta n$ as the tuning parameter.
\ Fifty four values of $\Delta n$ between 0 and 3.35 x 10$^{13}$ cm$^{-2}$
were included. \ The best collapse of the data was for $\nu z=0.7$ with an
uncertainty of $\pm0.1$. \ Inset: $R(\Delta n)$ for isotherms between 60 mK
and 140 mK.}%
\end{center}
\end{figure}
\ The value of the critical exponent product $\nu z$ that brings about the
best collapse of the data is $0.7\pm0.1$. The range over which scaling is
successful is from 60 mK to 140 mK.

The scaling analysis with uncorrected data appears to fail at temperatures
above 140 mK because of the $lnT$ dependence of the conductance in the normal
state.\ It is possible to successfully extend the analysis up to 1 K if this
dependence is first removed. This is done by assuming that there are two
parallel conductance channels, one that gives the $lnT$ dependence and the
other that gives superconducting and insulating fluctuations. We then find the
critical charge transfer to be at $\Delta n_{c}=$ 0.85 x 10$^{13}$ cm$^{-2}$
since at this temperature $R(T)$ can be fit by $lnT$ over the entire
temperature range (from 60 mK to 1 K). The best fit to $G(T)$ at $\Delta
n_{c}$ is then written as $G_{0c}+k_{c}lnT$ and the critical resistance is the
temperature independent part of the resistance at $\Delta n_{c}$, or
$1/G_{0c}$. For each $\Delta n$, we subtracted $k_{c}lnT$ from $G(T)$. After
this, we determined the value of $\upsilon z$ that minimized the error in the
data collapse on the scaling curves in the same manner as for the uncorrected
\ data. All values of $R(T)$ from 60 mK to 1 K then successfully fell onto the
scaling curves, as shown in Fig. 4.%
\begin{figure}
[ptb]
\begin{center}
\includegraphics[
height=2.0695in,
width=3.0199in
]%
{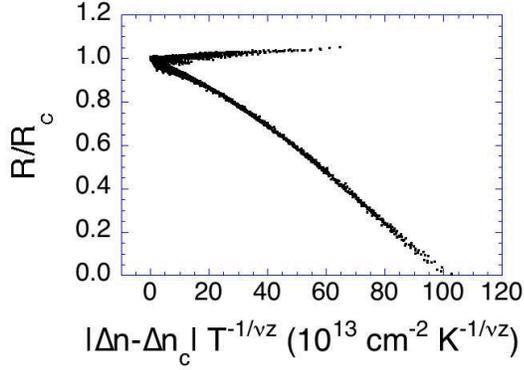}%
\caption{Finite size scaling plot for the 10.22 \AA \ thick film with $B=0$
including values of $R(T)$ from 60 mK to 1 K with $\Delta n$ as a tuning
parameter, having first removed the contributions to $R(T)$ from from the
conductance channel responsible for the $lnT$ dependence. The best collapse of
the data was for $\nu z=0.7$ with an uncertainty of $\pm0.1$.}%
\end{center}
\end{figure}
\ The critical exponent product remained at $0.7\pm\ 0.1$. The parallel
conductance channel responsible for the $ln(T)$ dependence is thus removed
from the analysis using this procedure. This approach is more physical than
that used by Gantmakher \textit{et al}. \cite{Gantmakher scaling} in the study
of perpendicular magnetic field tuned SI transitions of In$_{2}$O$_{3}$ films,
where scaling was successful at the lowest temperatures, but a positive
$dR_{c}/dT$ at high temperatures prevented scaling. To broaden the temperature
range they subtracted a linear temperature dependence from $R_{c}(T)$.

This SI transition appears to involve little change in the physical disorder,
as determined by noting that the resistance at high temperatures ($T\geq1$K)
changes very little as superconductivity develops. This resistance is related
to the product of the Fermi wave vector, $k_{F}$, and the electronic mean free
path$,l$. \ If $k_{F}l$ does not change much, then one can assume that
disorder does not change much. In Fig. 5,%
\begin{figure}
[ptb]
\begin{center}
\includegraphics[
height=2.0695in,
width=2.5607in
]%
{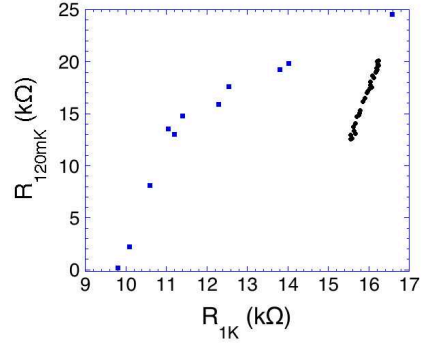}%
\caption{Resistance at 1 K as a function of resistance at 120 mK for
electrostatic (circles) and thickness (squares) tuned transitions. As
superconductivity develops, the resistance at 1 K changes much more for
thickness tuning than for electrostatic tuning.}%
\end{center}
\end{figure}
we show the resistance at low temperature (120 mK) as a function of the
resistance at 1 K for the electrostatic tuned transition and a previously
reported thickness-tuned transition.\cite{Parendo PRB}\ One can see that, as
insulating behavior changes to superconducting behavior, the thickness-tuned
transition takes place with a much larger change in the high temperature
resistance than does the electrostatically tuned system. To further the idea
that electrostatic doping adjusts electronic properties but not disorder, we
note that for intrinsically superconducting films with transition temperatures
between 1 and 4 K, the application of $V_{G}=$ \ $\pm50$ V shifted transition
temperatures by as much as 40 mK, but these shifts were accompanied by very
small changes in the resistance at 10 K. The later changed by no more than
$10-20$ $\Omega$ with normal state resistances of 3-5 k$\Omega$.

\subsection{Electrostatic Tuning of the SI Transition in a 2.5 T Parallel
Magnetic Field}

Electrostatic tuning of the SI transition of the 10.22 \AA \ thick film was
also carried out in a parallel magnetic field of 2.5 T. The data is shown in
Fig. 6.%
\begin{figure}
[ptb]
\begin{center}
\includegraphics[
height=2.0695in,
width=2.2572in
]%
{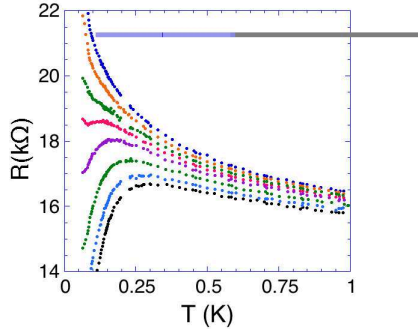}%
\caption{$R(T)$ as a function of $\Delta n$ for the 10.22 \AA \ thick film
with $B=2.5$ T. Values of $\Delta n$ shown are 0, 0.74, 1.28, 1.57, 1.87,
2.30, 2.74 and 3.13 x 10$^{13}$cm$^{-2}$, from top to bottom. Fifteen curves
of $R(T)$ at other values of $\Delta n$ have been omitted from the plot for
clarity.}%
\end{center}
\end{figure}
\ The application of magnetic field to an ungated insulating film increased
its activation energy. Adding electrons increased the film's conductivity and,
at high enough density, induced a transition to superconductivity.

Comparing Figs. 2 and 6, a major difference between the electrostatically
tuned SI transitions in zero and non-zero magnetic fields is evident: at low
temperatures, in nonzero field, a regime exists in which the resistance at the
lowest temperatures is larger than that expected from the extrapolation of
$R(T)$ from higher temperatures. This is most obvious for electron densities
around the critical density, where curves that appear to be heading towards
zero resistance as temperature is lowered suddenly undergo a change in slope
and appear to be insulating when extrapolated to the limit of zero temperature.

A finite size scaling analysis was carried out for the electrostatically tuned
transition in field.\ In the inset to Fig. 7 we show $R(\Delta n)$ for
isotherms between 100 and 200 mK.%
\begin{figure}
[ptb]
\begin{center}
\includegraphics[
height=2.0695in,
width=3.0199in
]%
{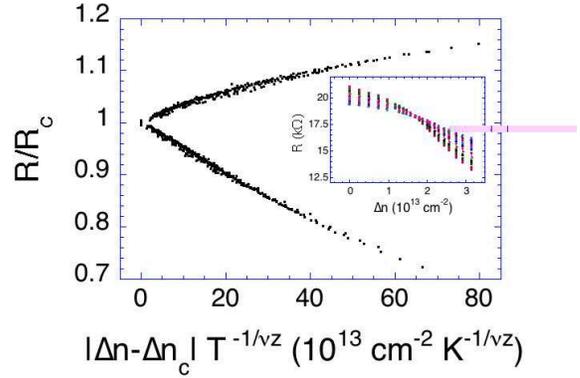}%
\caption{Finite size scaling plot for the 10.22 \AA \ thick film with $B=2.5$
T including values of $R(T)$ from 100 mK to 200 mK with $\Delta n$ as the
tuning parameter. \ Twenty three values of $\Delta n$ between 0 and 3.13 x
10$^{13}$ cm$^{-2}$ were included. \ The best collapse of the data was for
$\nu z=$ 0.65 with an uncertainty of $\pm$0.1. \ Inset: $R(\Delta n)$ for
isotherms between 100 mK and 200 mK, exhibiting a well-defined crossing point
at $\Delta n=1.7$ x10$^{13}$cm$^{-2}$ and $R=18,$300 $\Omega.$}%
\end{center}
\end{figure}
\ A distinct crossing point is apparent, yielding $\Delta n_{c}=$ 1.7 x
10$^{13}$ cm$^{-2}$ and $R_{c}=$ 18,300 $\Omega$. In the main body of Fig. 7,
we show the scaling plot. The value of the exponent product that minimized the
collapse of the data was $0.65\pm0.1$, which is slightly lower than the
electrostatic tuned transition in zero field but is in agreement with it if
the uncertainty of the analysis is taken into account. Data below 100 mK does
not fall onto the scaling plot because of the above mentioned excess resistance.

The successful scaling analysis implies that the system is moving towards
insulating and superconducting ground states that are separated by a quantum
critical point. The break down in scaling implies that instead of reaching
this insulating ground state, one with a higher resistance is actually
achieved. This will be discussed in more detail in in Section F.

\subsection{Parallel Magnetic Field Tuning of the SI Transition}

In the case of the 10.22 \AA \ thick film, after superconductivity was induced
by adding an areal density of carriers $\Delta n=3.35$ x 10$^{13}$ cm$^{-2}$,
this induced superconductivity was quenched by a parallel magnetic field, and
a detailed study of the parallel magnetic field tuned SI transition was
conducted.\ In Fig. 8, we show this transition.%
\begin{figure}
[ptb]
\begin{center}
\includegraphics[
height=2.0695in,
width=2.1248in
]%
{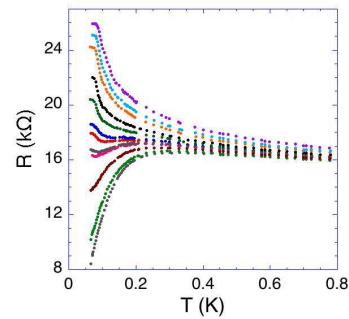}%
\caption{$R(T)$ as a function of B for an insulating film with
superconductivity induced by a charge transfer $\Delta n=3.35$ x
10$^{13}cm^{-2}$. \ Values of $B$ are 2 (bottom), 2.5, 3.5, 4.25, 4.375, 4.75,
5, 5.75, 6.5, 8, 9, and 11 T (top). \ Seven $R(T)$ curves for other values of
B have been omitted from the plot for clarity.}%
\end{center}
\end{figure}
\ With B = 0, the superconducting transition temperature, $T_{c}=60$ mK (taken
to be the highest temperature at which resistance is zero within the scatter
due to noise), the mean field transition temperature (taken to be the
temperature at which resistance is half of the normal state peak) is 90 mK,
and the peak in $R(T)$ below which there are strong superconducting
fluctuations is at 250 mK.\ The normal state, taken to be at temperatures in
excess of that at which $R(T)$ exhibits a peak, is best described by a
conductance that has a logarithmic temperature dependence. This superconductor
at $B=0$ becomes an insulator at high fields that is again well described by
2D Mott VRH at temperatures greater than 130 mK. \ 

Analysis of this transition using finite size scaling is successful over a
wide range of temperature, provided that the regime of excess resistance
discussed above is excluded from the analysis. The scaling analysis is shown
in Fig. 9.%
\begin{figure}
[ptb]
\begin{center}
\includegraphics[
height=2.0695in,
width=3.0199in
]%
{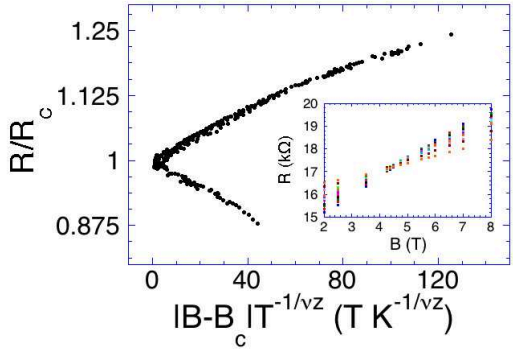}%
\caption{Finite size scaling plot for the 10.22 \AA \ thick film with $\Delta
n=3.35$x 10$^{13}$ cm$^{-2}$ with $B$ as the tuning parameter for 150 mK $<T<$
340 mK. The best collapse of the data was for $\nu z=$ 0.65 with an
uncertainty of $\pm$0.1. \ Inset: Isotherms of $R(B)$ at temperatures between
150 mK and 340 mK.}%
\end{center}
\end{figure}
\ The inset to Fig. 9 shows a set of isotherms of resistance vs. field between
150 and 340 mK, which exhibits a distinct crossing point at $R_{c}=17,285$
$\Omega$ and $B_{c}=4.625$ T. \ The value of the exponent product $\nu z$ that
minimized collapse of the data is $0.65\pm0.1$. Within the experimental
uncertainty, this value is the same as that found for the electrostatically
tuned transitions in zero field and 2.5 T.

\subsection{Phase Diagrams}

We have carried out a systematic study of the parallel magnetic field tuned SI
transition at various strengths of electrostatically induced
superconductivity. The scaling procedure for the B-tuned transition for
$\Delta n=3.35$ x 10$^{13}$ cm$^{-2}$ discussed in the previous section was
repeated at $\Delta n=$1.66, 2.25, and 2.80 x 10$^{13}$ cm$^{-2}$, yielding
the same exponent product $\nu z=$ $0.65\pm0.1$.

At values of $\Delta n=$ 1.83, 2.00, 2.40, 2.59 and 2.96 x 10$^{13}$ cm$^{-2}%
$, $R(B)$ was measured at 100 and 120 mK and the crossing points determined.
These temperatures were within the range used to determine the crossing points
when a full scaling analysis was carried for $\Delta n=$ 1.66, 2.25, and 2.80
x 10$^{13}$ cm$^{-2}$ using data at many different temperatures. Thus, they
are greater than the temperatures at which excess resistance occurs. Because
the crossing points were well defined when there was extensive data, one can
be confident that the intersection at two temperatures would be a reliable
determination of the critical resistance and critical field. \ 

Using both the full scaling analyses and the reduced crossing-point analyses,
we were able to map out the variations of both $B_{c}$ and $R_{c}$ with
$\Delta n$ as shown in Figs. 10 and 11, respectively. The critical field,
$B_{c}$, increased with $\Delta n$, as $(\Delta n-\Delta n_{c})^{0.33}$.%
\begin{figure}
[ptb]
\begin{center}
\includegraphics[
height=2.0695in,
width=2.9144in
]%
{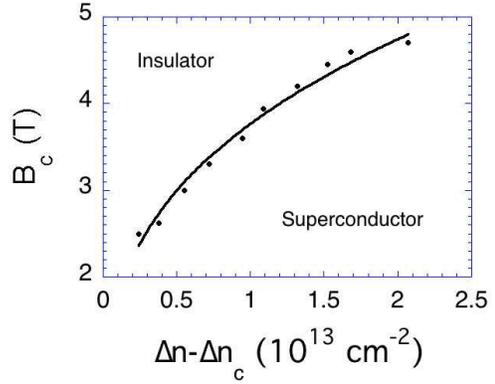}%
\caption{Phase diagram of $B_{c}$ vs. $(\Delta n-\Delta n_{c})$ for the 10.22
\AA \ thick film. \ The best fit is by a power law with an exponent of 0.33. }%
\end{center}
\end{figure}
\ Qualitatively, this is what would be expected, since adding carriers
strengthens superconductivity and destroying stronger superconductors would
require higher magnetic fields. The critical resistance decreased linearly
with $\Delta n$. The data from the electrostatically tuned transition in a
magnetic field of 2.5 T is included on these plots. \ If this data is
excluded, the functional forms are unchanged. \ 

The exponent product $\nu z$ is the same, within experimental uncertainty, for%
\begin{figure}
[ptb]
\begin{center}
\includegraphics[
height=2.0695in,
width=3.0199in
]%
{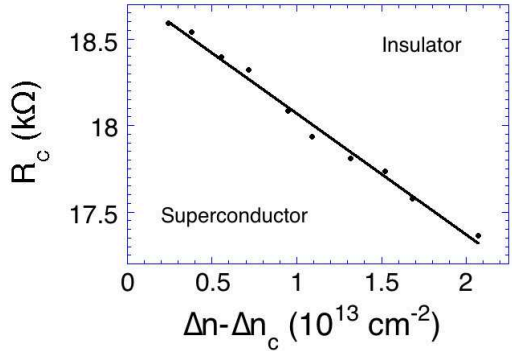}%
\caption{Phase diagram of $R_{c}$ vs. $(\Delta n-\Delta n_{c}$) for the 10.22
\AA \ thick film. \ The best description is a linear relation.}%
\end{center}
\end{figure}
the parallel magnetic field tuned transitions at various strengths of
electrostatically induced superconductivity and the electrostatically-tuned
transitions in zero and finite magnetic fields. This suggests that the quantum
phase transitions belong to the same universality class and that these phase
diagrams demark the insulating and superconducting regimes with a robust phase
transition line that can be crossed in either direction, by tuning $B$ or
$\Delta n$. This result is differs from that found by Markovic \textit{et
al}.\cite{Markovic Phase} in which the exponent products of thickness and
perpendicular field tuned transitions were found to be different. Note that
the regime of excess resistance found in parallel magnetic fields may imply
that the insulating ground state may be different from that which might be
deduced from the analysis of data obtained at high temperatures and used to
construct Figs. 11 and 12.\ \ 

\subsection{Excess Resistance in High Magnetic Fields}

In films whose parameters place them on the border between the insulating and
superconducting regimes, or well into the insulating regime, we observed
behavior in which resistance in magnetic fields became higher at low
temperatures than what might have been expected from extrapolation of R(T) at
higher temperatures. This regime appears to involve physics that is different
from that which determines quantum criticality. While a scaling analysis is
successful including data obtained at higher temperatures, it fails when data
from this regime is included. \ 

To illustrate this excess resistance, we now discuss in more detail the data
on the parallel magnetic field-tuned transition shown in Fig. 9. Here, the
resistance is larger than what is expected by extrapolating the curves of
$R(T)$ down from temperatures in excess of 150 mK. \ Between 3 T and the
parallel critical field of 4.625 T, while superconducting fluctuations cause
$R(T)$ curves to head towards zero resistance as temperature is lowered, there
is a small upturn in resistance as 60 mK is approached. This upturn in $R(T)$
is found at all higher fields in the insulating regime. It is easily visible
at fields between 4.625 T and 7 T, where the upturn occurs simultaneously with
a minimum in $R(T)$. In fields in excess of 7 T, there are no minima in $R(T)$
since the general shape is that of an insulating curve with $dR/dT<0$ for all
$T$, but there is still an excess resistance at low temperatures.\ Here,
resistance is in excess of an extrapolation of Mott VRH, which describes the
data from 130 mK up to 1 K. To illustrate this, $R(T)$ \ in fields of 8, 9,
and 11 T from Fig. 8 are re-plotted in Fig. 12 in the%
\begin{figure}
[ptb]
\begin{center}
\includegraphics[
height=2.0695in,
width=3.0199in
]%
{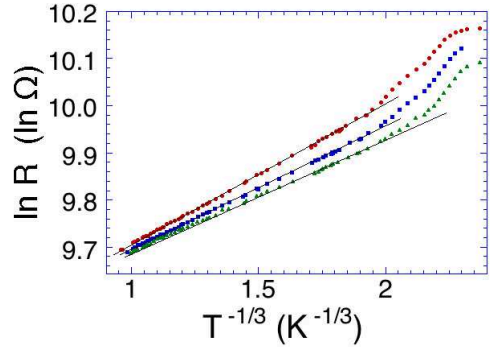}%
\caption{With superconductivity first induced in the 10.22 \AA \ thick film
with $\Delta n=$ 3.35 x 10$^{13}$cm$^{-2}$, $ln(R)$ vs. $T^{-1/3}$ for
insulating curves induced by 8 (bottom), 9, and 11 (top) T. The data is fit
well by Mott VRH for $T>$150 mK, but deviates to a higher resistance below 150
mK. \ Straight lines have been added as guides to the eye.}%
\end{center}
\end{figure}
form $ln(R)$ vs. $T^{-1/3}$.\ At higher temperatures, 2D Mott VRH is seen to
be a good fit, while below about 130 mK, the resistance becomes greater than
extrapolations using the Mott form from higher temperatures. \ 

The minima in $R(T)$ and the increase in resistance at temperatures below
those at which the minima occur are both qualitatively similar to features in
granular superconducting films. In granular films, minima in $R(T)$ and the
subsequent increase of resistance with decreasing temperature are believed to
be indicative of persistence of superconductivity on mesoscopic sized grains,
where inter-granular transport is via single-particle tunneling. We have not
observed reentrance in the absence of magnetic field, either in the
electrostatically tuned SI transition or the thickness-tuned transition (see
Figs. 1 and 2). This suggests that there are no mesoscopic scale clusters in
the film.

The magnitude of the excess resistance in magnetic field in these films is
much smaller than that in granular films. For instance, for $B=$ 8, 9, and 11
T, resistances become about 4-8 \% higher at 65 mK than one would expect from
extrapolation. At these fields, the minima occur at about 140 mK. In contrast,
in granular films at temperatures a factor of two lower than the temperature
at a minimum, resistance is found to be some 16 times higher than is
predicted, due to a $(T/T_{min})^{4}$ dependence associated with the opening
of the energy gap as superconductivity develops on the grains.\cite{Adkins,
Eytan}

The highest temperature at which excess resistance is observed in a given
field, denoted as $T_{onset}$, increases with increasing field. This was
checked for the four sets of parallel field induced SI transitions for which
full scaling analyses were performed, as well as for an intrinsically
insulating film where superconductivity was not induced
electrostatically.\ For each superconducting film, the onset of excess
resistance appears around the critical magnetic field of the SI transition.
\ As mentioned before, $B_{c}$ increases with $\Delta n$.\ Interestingly,
these $T_{onset}(B)$ curves collapse if shifted horizontally to align the
critical fields.\ This is shown in Fig. 13.%
\begin{figure}
[ptb]
\begin{center}
\includegraphics[
height=2.0695in,
width=3.0199in
]%
{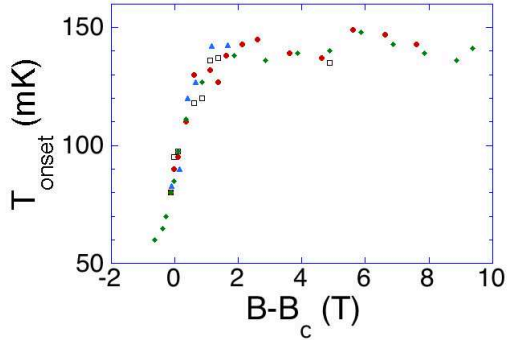}%
\caption{$T_{onset}$ vs. $B-B_{c}$ for superconducting films with various
values of $\Delta n$. $\Delta n=1.66$ (diamonds), 2.25 (triangles), 2.80
(squares), and 3.35 (circles) x $10{{}^1}{{}^3}$ cm$^{-2}$. The data has been
collapsed by shifting these curves by 2, 2.7, 3.5, and 3.75 T, respectively,
from the unadjusted values.}%
\end{center}
\end{figure}
\ This implies that when superconductivity is stronger (at larger $\Delta n$)
it takes a larger field to induce excess resistance than it takes when
superconductivity is weaker (at smaller $\Delta n$).\ It also implies that
once excess resistance has been induced, it develops with increasing field in
a manner that is independent of the initial strength of superconductivity.

There have been several observations of a peak in the magnetoresistance at
fields in excess of the critical field for the SI transition in thicker films
($\backsim50-300$ \AA ) of In$_{2}$O$_{3}$ and of TiN, both in fields aligned
perpendicular\cite{Paal Heb Ruel, Gantmakher metal, Villegier, Shahar, Steiner
and Kapitulnik, Baturina} to and parallel\cite{Gantmakher parallel} to the
film plane. This peak has even been observed in insulating Be
films.\cite{Butko and Adams} \ Typically, this resistance peak is found to be
larger in films with high normal state resistances (and correspondingly low
values of $T_{c}$) than in those with low normal state resistances and higher
values of $T_{c}$.\cite{Steiner and Kapitulnik}\ When we induce
superconductivity in the 10.22 \AA \ \ thick film with $\Delta n=$ 3.35 x
10$^{13}$ cm.$^{-2}$, with $R_{N}\backsim16$ k$\Omega$ and $T_{c}\backsim60$
mK, we do not observe a peak in resistance as a function of field.\ In Fig.
14, we show R(B) for various isotherms between 65 mK and 200 mK.%
\begin{figure}
[ptb]
\begin{center}
\includegraphics[
height=2.0695in,
width=2.6991in
]%
{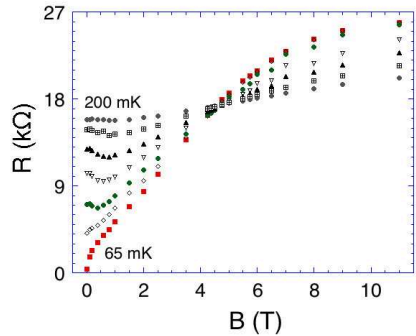}%
\caption{$R(B)$ for the 10.22 \AA \ thick film with $\Delta n=$ 3.35 x
10$^{13}$ cm$^{-2}$. \ The isotherms correspond to 65, 75, 85, 100, 120, 150,
and 200 mK. \ At high fields, $dR/dB>0$.}%
\end{center}
\end{figure}
\ For all isotherms, $dR/dB>0$ at high fields.\ This appears to be a general
feature of our films, as we observed $dR/dB>0$ up to 12 T at all temperatures
in the other parallel magnetic field tuned transitions at smaller values of
$\Delta n$.

\subsection{Comparison of the Field-Tuned SI Transitions of Intrinsic and
Electric-Field Induced Superconductors}

A separate experiment on parallel field tuning of the SI transition was
conducted on an intrinsically superconducting film.\ The results of a finite
size scaling analysis were nearly identical to those found for the transition
of an electrostatically-induced superconducting film. The excess resistance in
high fields was also nearly identical to that found in an
electrostatically-induced superconducting film. The sample was a 10.25
\AA \ thick\textit{ a-}Bi film deposited on top of 10 \AA \ thick\textit{
a-}Sb layer on a STO substrate that had not been thinned.\ This film was
superconducting with a transition temperature $T_{c}=130$ mK and a normal
state sheet resistance, $R_{N}$ $\backsim9500$ $\Omega$.\ Parallel magnetic
fields induced insulating behavior and finite size scaling was successful over
a range of temperatures from 150 to 400 mK. A distinct crossing point was
found, with $R_{c}=10,460$ $\Omega$ \ and $B_{c}=$ 9.18 T. The critical
exponent product that produced the best collapse of the data was $0.75\pm0.1$,
slightly higher than found for the field-tuned transition of an
electrostatically induced superconducting film, but agreeing within the
uncertainty. Excess resistance was observed in fields ranging from 8 T to 12 T
at temperatures below about 150 mK. \ At 12 T, a fit to $R(T)$ by Mott VRH was
successful down to 150 mK. For data below 150 mK, the excess resistance again
prevented successful scaling. \ For T
$>$
400 mK, the negative slope of $dR_{c}/dT$ prevented successful scaling. At
high fields and at all temperatures, $dR/dB$ was always positive.

\section{Discussion}

\subsection{Magnetic Field Alignment}

An important technical issue for studies in parallel magnetic fields is the
alignment of the plane of the substrate with the field. Our estimate of the
maximum angular misalignment of the is 1$^{0}$.\ This is based on multiple
geometrical constraints. For the magnetic field tuned SI transition for
superconductivity induced by adding $\Delta n=$ 3.35 x 10$^{13}$ cm$^{-2}$,
the transition temperature is 60 mK, the "mean field" transition temperature
is 90 mK, and the critical field is 4.625 T. This critical field is 27 times
higher than the limiting value necessary to destroy superconductivity by
aligning spins, found by Clogston and Chandrasekhar to be 1.9 T/K. An
enhancement of this magnitude might be expected as Bi is a heavy metal with
strong spin-orbit interaction and consequently\textit{ a-}Bi would be expected
to have a short \ spin orbit scattering time. At each electron scattering
event, the spin flips, leading to a substantially enhanced value of the
critical field relative to the Clogston/Chandrasekhar limiting field. If we
were to assume that this result was due to a perpendicular field component
resulting from misalignment larger than our estimate, a misalignment of about
7 to 11 degrees would be necessary to produce the critical fields in Fig. 10.
We base this on comparison with work on similar \textit{a-}Bi films by
Markovic\textit{ et al}.\cite{Nina 2D SIT} in which a perpendicular field of
0.6 T was needed to quench superconductivity in a film that has a transition
temperature between that found here at $\Delta n=$ 1 x 10$^{13}$ cm$^{-2}$ and
3.35 x 10$^{13}$ cm$^{-2}$, based on its shape. \ This is unreasonably large,
given the geometrical constraints.

\subsection{SI Transitions}

In aggregate, finite size scaling analyses of the various SI transitions tuned
by thickness, electrostatic electron doping, and perpendicular and parallel
magnetic fields have yielded a wide range of \ results. Scaling analyses of
measurements in which perpendicular magnetic field,\cite{Yazdani and
Kapitulnik, Hebard and Paalanen, Gantmakher parallel, Gantmakher metal,
Bielejec and Wu, Nina 2D SIT} or thickness\cite{Markovic Phase} have been
tuned have usually resulted in $\nu z\backsim1.3$ and $z=1$, which is
consistent both with the scaling theory,\cite{FGG} (2+1) dimensional XY model
with disorder,\cite{Cha and Girvin} and other and numerical
studies.\cite{Makivic, CRGWY, SWGY, Singh, Zhang, Herbut} \ This value has
been suggested to be consistent with percolation.\cite{Shimshoni and
Kapitulnik, Meir} An exception is the case of \textit{a-}Bi, in which tuning
the transition by a perpendicular field has yielded $z\backsim0.7$, while
tuning with thickness has yielded $z$ $\backsim1.3$.\cite{Nina 2D SIT}
Assuming that $z=$ 1, this value disagrees with what was believed to be an
theorem that predicts $\nu\geq1$ in two dimensions in the presence of
disorder.\cite{Chayes} The possible irrelevance of this theorem to the SI
transition has been discussed most recently by Chamon and Nayak.\cite{Chamon}
The present measurements suggest that transitions tuned by electron doping and
parallel magnetic fields have exponent products of 0.65 to 0.75. Thus if
$z=1$, $\nu z=$ 0.7 is consistent with earlier perpendicular field scaling, as
well as numerical studies of the 3D XY model\cite{FWGF, Cha and Girvin} and
the Boson-Hubbard models in the absence of disorder.\cite{Kisker} This is
suggestive that thickness-tuned transitions in \textit{a-}Bi may be
percolative, while transitions for other tuning parameters are not. \ 

The values of critical resistances for all of the transitions of
electrostatically induced superconductivity are a factor or two or three
higher than those found in other SI transitions and the universal value of
6455 $\Omega$ predicted by the dirty Boson model.While the critical resistance
has been found experimentally to be non-universal,\cite{Yazdani and
Kapitulnik, Nina 2D SIT} it has been studied in various materials and samples
with different levels of disorder. Here, in a sample with a static level of
disorder, we have shown the critical resistance decreases linearly with
increasing electron density. \ 

Excess resistance appears to be a low temperature feature of films in the
insulating regime in the presence of parallel magnetic fields. Fields have
produced this both insulating regime when superconductivity was quenched by
field and in the intrinsic insulating state. The features of $R(T)$ in
parallel magnetic fields are qualitatively similar to features in $R(T)$ in
zero field for granular films. The conventional view is that quench-condensed
films grown with either an \textit{a-}Ge or an \textit{a-}Sb underlayers are
disordered on an atomic rather than a mesoscopic scale. Since these films
anneal at temperatures around 25 K, whereupon they change structurally into
semimetals, we cannot examine them structurally to determine the level of
homogeneity. However, since we have not observed reentrance in the absence of
magnetic field, either in the electrostatically tuned SI transition or the
thickness-tuned transition, we believe that there are no mesoscopic scale
clusters in the film.\ It is possible that the parallel magnetic field induces
spatial inhomogeneity of the amplitude of the superconducting order parameter
that mimics clustering.

In the present work, we found that the relation between critical parallel
magnetic field and electron density is $B_{c}\backsim(\Delta n-\Delta
n_{c})^{0.33}$, whereas Markovic\textit{ et al}.\cite{Markovic Phase} found
the relationship between perpendicular magnetic field and film thickness, d,
as $B_{c}\backsim(d-d_{c})^{0.7}$ for perpendicular fields. There appears to
be no theory constraining the relationship between $\Delta n$ and $B_{c}$. It
appears that the line of criticality can be crossed changing either $B$ or
$\Delta n$ with a finite size scaling analysis yielding the same exponent
product, whereas in the work of Markovic \textit{et al}. tuning with $d$ and
with $B$ produced different exponent products.

Flattening of $R\left(  T\right)  $ below about 60 mK in zero magnetic field
may only be a consequence of failure to cool the film. In nonzero magnetic
field its occurrence \ at temperatures higher than 60 mK may be due to
enhanced Eddy current heating or there may be an intrinsic metallic regime.

Interestingly, we have not observed a peak in the magnetoresistance after
superconductivity has been quenched by magnetic fields, even though our
transition temperatures are very low and normal state resistances are very
high, which is the regime in which the largest peaks of In$_{2}$O$_{3}$ films
were found.\cite{Steiner and Kapitulnik} \ We do not find a peak either below
150 mK, in the regime of excess resistance, or above this temperature, where
Mott VRH fits well. \ Though our fields are aligned parallel to the plane of
the film, large peaks in magnetoresistance were found by Gantmakher et al. in
alignments both perpendicular and parallel to the film.\cite{Gantmakher metal,
Gantmakher parallel}\ Though there is theoretical work suggesting that this
metallic regime is intrinsic,\cite{Fisher 2005} there have been suggestions
that this is a consequence of inhomogeneity in the films.\cite{Shahar,
Lopatin} We suggest that further experimental studies of the homogeneity of
films exhibiting peaks in magnetoresistance in the presence of large magnetic
fields would help resolve the issue of whether these magnetoresistance peaks
are intrinsic properties of structurally and chemically homogeneous films.

\section{Conclusions}

We have investigated the two-dimensional superconductor-insulator transition
in disordered ultrathin \textit{a-}Bi films by use of electrostatic electron
doping using the electric field effect and by the use of parallel magnetic
fields. Electrostatic doping was carried out in both zero and nonzero magnetic
fields, and magnetic tuning was conducted at multiple strengths of
electrostatically induced superconductivity.\ The various transitions were
analyzed using finite size scaling to determine the critical exponent products
of the quantum phase transitions, which were all found to be $\nu z=$ 0.65 or
$0.7\pm0.1$. The critical parallel magnetic field increased with electron
transfer, $\Delta n$, as $(\Delta n_{c}-\Delta n)^{0.33}$, while the critical
resistance decreased linearly with electron transfer. \ 

An anomalous regime of excess resistance that is induced by a parallel
magnetic field was also observed. This excess resistance was not observed in
zero magnetic field for either thickness or electrostatic tuning. the
existence of this regime necessitated that data below 100 to 150 mK be
excluded in order for scaling to be successful.

Although there is a long history of experimental and theoretical investigation
of the control of superconductivity with electric fields beginning with Glover
and Sherill,\cite{Glover} there has been no significant tuning of
superconductivity in metallic systems, and models based on the BCS
theory\cite{Lipavsky, Shapiro, Lee} only treat the effect on the transition
temperature of changes in the density of states in response to changes in the
carrier concentration. They do not include an issue that is relevant here, the
apparent insulator to metal transition that accompanies the superconductor to
insulator transition.\cite{Parendo PRL} The change in the coefficient of the
$lnT$ behavior of the resistance in the normal state reported in
Parendo\textit{ et al}.\cite{Parendo PRL} implies that the effective
electron-electron interaction is also changed as the carrier concentration
(electron density) is increased. This is not included in any of the
theoretical treatments and is a challenge for theory. \ 

The parallel field tuned SI transition leads to values of the critical field
that are the order of a factor of 30 in excess of the Pauli limiting field for
a superconductor with a transition temperature of 60 mK. This is to be
expected for a system with strong spin-orbit scattering. \ In this instance
the rather extensive theoretical treatments\cite{Fulde} of the critical fields
of superconductors are not applicable as quantum fluctuations and the fact
that the normal phase is an insulator and not a metal have not been taken into account.

Finally the excess resistance in the insulating phase of the parallel field
tuned transition, which actually appears like quasi-rentrant
superconductivity, strongly suggests the existence of a regime in which the
superconducting order parameter is inhomogeneous, i.e., that there may be
superconducting droplets in an insulating matrix. \ Although this was observed
only in the field-tuned transition and not in the charge-tuned transition, the
idea is fairly generic in theories of SI transitions in homogeneously
disordered films.\ The case of perpendicular fields has been considered by
Spivak and Zhou\cite{Spivak} and by Galitski and Larkin\cite{Larkin and
Galitski}. For thin films with strong spin-orbit scattering, as would be the
case for \textit{a-}Bi, Zhou\cite{Zhou} has shown that there should be a
glassy phase. We have observed very slow relaxation in hgih magnetic fields,
which could be evidence of such a reigme. In all instances the inhomogeneous
regime is confined to a narrow region around the critical field. Recently
Skvortsov and Feigelman\cite{Skvortsov} have considered the suppression of
superconductivity by disorder in films of high dimensionless conductance. They
find an inhomogenous regime close to the critical conductance where
superconductivity is completely suppressed. This is not seen in our
electrostatically tuned transition, perhaps due to the relatively low normal
state conductance in our films.

\begin{acknowledgments}
\bigskip The authors are grateful to the late Anatoly Larkin for numerous
discussions and dedicate this work to him.\ They would like to thank A. Efros,
B. Shklovskii, L. Glazman, A. Kamenev, P. Crowell, J. Meyer, A. Finkel'stein,
and Y. Imry for many helpful discussions. They would also like to thank A.
Bhattacharya, M. Eblen-Zayas, and N. Staley for their work with SrTiO$_{3}$
substrates that facilitated this research. This research was supported by the
National Science Foundation under grant NSF/DMR-0455121.
\end{acknowledgments}

\bigskip

\end{document}